%% file: ms.tex
\documentclass[]{spie} 
\usepackage[]{graphicx}
\usepackage{wrapfig,sidecap,subcaption}
\usepackage{amssymb,amsmath,wasysym,upgreek}
\usepackage{color}
\input{mymacros}

\title{Characterization of a photon counting EMCCD for space-based high contrast imaging spectroscopy of extrasolar planets} 

\author{Ashlee N. Wilkins\supit{a}, Michael W. McElwain\supit{b}, Timothy J. Norton\supit{b,c}, Bernard J. Rauscher\supit{b}, Johannes F. Rothe\supit{d}, Michael Malatesta\supit{e}, George M. Hilton\supit{a,b}, James R. Bubeck\supit{b,f},Carol A. Grady\supit{b,g}, Don J. Lindler\supit{b}
\skiplinehalf
\supit{a}University of Maryland, College Park, MD, USA;
\skiplinehalf
\supit{b}NASA Goddard Space Flight Center, Greenbelt, MD, USA;
\supit{c}University of Maryland, Baltimore County, Baltimore, MD, USA;
\supit{d}Universities Space Research Association, Columbia, MD, USA; 
\supit{e}University of Oklahoma, Norman, OK, USA;
\supit{f}Adnet Systems, Rockville, MD, USA;
\supit{g}Eureka Scientific, CA, USA;
}

\authorinfo{Send correspondence to A.N.W.\\A.N.W.: E-mail: ashlee.n.wilkins@nasa.gov, Telephone: (301) 286-9121 \\}

\begin{document} 
 \maketitle 

\begin{abstract}
We present the progress of characterization of a low-noise, photon counting Electron Multiplying Charged Coupled Device (EMCCD) operating in optical wavelengths and demonstrate possible solutions to the problems of Clock-Induced Charge (CIC) and other trapped charge through sub-bandgap illumination. Such a detector will be vital to the feasibility of future space-based direct imaging and spectroscopy missions for exoplanet characterization, and is scheduled to fly on-board the AFTA-WFIRST mission. The 512$\times$512 EMCCD is an e2v detector housed and clocked by a N\"uv\"u Cameras controller. Through a multiplication gain register, this detector produces as many as 5000 electrons for a single, incident-photon-induced photoelectron produced in the detector, enabling single photon counting operation with read noise and dark current orders of magnitude below that of standard CCDs. With the extremely high contrasts (Earth-to-Sun flux ratio is $\sim$ 10$^{-10}$) and extremely faint targets (an Earth analog would measure 28$^{th}$ - 30$^{th}$ magnitude or fainter), a photon-counting EMCCD is absolutely necessary to measure the signatures of habitability on an Earth-like exoplanet within the timescale of a mission's lifetime, and we discuss the concept of operations for an EMCCD making such measurements. 

\end{abstract}

\keywords{photon counting detector,, electron multiplying CCD high contrast, exoplanets, direct imaging, integral field spectroscopy, clock induced charge, cameras, clocks, exoplanet spectroscopy}

\section{INTRODUCTION}
\label{sec:intro} 
Photon counting, electron-multiplying CCDs (EMCCDs) are under strong consideration for missions as imminent as the AFTA-Coronagraph (technology deadline 2017)\cite{Shaklan2013}, and larger missions further in the future, like the Advanced Technology Large Aperture Space Telescope (ATLAST)\cite{Postman2010}. This technology will be matured as part of the Prototype Imaging Spectrograph for Coronagraphic Exoplanet Studies\cite{McElwain2013} IFS to be installed at the High Contrast Imaging Testbed\cite{Trauger2007} in 2015. EMCCDs are a highly desirable technology for exoplanet imaging and spectroscopy missions, as identified by the Astro2010 white paper\cite{Levine2009}. When performing starlight-suppressed, direct imaging/spectroscopy observations of Earth analogs, photon rates are extremely low (on the order of one photon per thousand seconds). Photon counting detectors significantly increase the efficiency of observing such systems, to the point that integration times fit significantly better into the lifespan of a typical space-based mission. 

EMCCDs operate just as the name suggests: they measure the signal of a single photon on the detector after passing the corresponding photoelectron through a multiplication gain register that increases the output by two to three orders of magnitude\cite{Hynecek2003}. EMCCD operation improves as the source photon flux gets lower (in direct contrast to standard CCD operation), approaching the photon noise limit. Exposure (single frame) times are very short in photon counting mode due to the overlapping probability densities of recording more than one photon on a given pixel. Photon counting operation via thresholding requires each pixel to report signal in binary: pixel values are simply zero or one. As the rate of incoming photons increases -- or the frame rate of reading the detector decreases -- the probability of receiving more than one photon in a given pixel increases. The information is lost, and thus the accuracy is decreased, in this situations known as `pulse pile-up' or 'coincidence losses', the detector does not distinguish between single-photon events and multi-photon events.

The gain register renders read noise irrelevant, but it amplifies image-layer noise not usually of concern in standard CCD operation, known as clock-induced-charge (CIC). The major limitations of EMCCDs include both CIC and dark current\cite{Wen2006}, which, similarly, becomes more relevant at low photon rates. Further, what is generally measured and called CIC, is likely a combination of actual image-layer CIC and ``CIC" generated within the EM gain register\cite{Tulloch2008}. Over the last decade, as the role for EMCCDs in astronomy has become more apparent and imminent, efforts to reduce the CIC (at least from the image area) have increased\cite{MacKay2004,Wen2006,Ives2008,Tulloch2010}. The lowest combined dark current and CIC background noise floor is claimed by the Canadian company, N\"uv\"u Cameras, with their HN\"u camera system housing e2v EMCCDs\cite{Daigle2009,Daigle2010,Daigle2012}. The specifications for the e2v detector housed and clocked by a N{\"u}v\"u camera are in Table~\ref{table:specs}. 

The lower that background noise floor, the more efficient an observing strategy can be adopted, as less time can be spent to achieve the same signal-to-noise, which we describe in detail and quantify in \S\ref{sec:science}. This will be especially important in upcoming planet-finding direct imaging missions. The work described in what follows supports this premise by first verifying the claims of significant improvement in both CIC and dark current in the N\"uv\"u system (\S\ref{sec:TechApproach}), and then seeking to improve the operations even further by applying sub-bandgap illumination to the detector. This method is an effort to mitigate the image-layer CIC by significantly shorten the lifetime of populated traps without producing unwanted signal within the dynamic range of the detector (\S\ref{sec:sub-bandgap}). 
 
\begin{table}
\begin{center}
 \begin{tabular}{| c | c |}
 \hline
 \multicolumn{2}{|c|}{\bf{Detector}}\\ \hline
 Company & e2v \\ \hline
 Model & CCD97 \\ \hline
 Area & 512 $\times$ 512 pixels\\ \hline
 Pixel Size & 16 $\times$ 16 $\mu$m\\ \hline
 Spectral Range & 250 - 1100 nm \\ \hline
 EM pixel well depth & 800 ke$^-$\\ \hline
 \multicolumn{2}{|c|}{\bf{Camera Housing}} \\ \hline
 Company & N{\"u}v{\"u} \\ \hline
 Model & Hn{\"u} 512 with low fringing \\ \hline
 EM Gain & 1 - 5000 \\ \hline
 Pixel Readout Rate & 0.1, 1, or 3.33 MHz (Conventional, Horizontal/Serial) \\
 & 0.1, 1, or 3.33 MHz (Conventional, Vertical/Parallel) \\ 
 & 1, 5, 10, or 20 MHz (EM, Horizontal) \\
 & 0.2, 1, 2, or 3.33 MHz (EM, Vertical) \\ \hline
 \end{tabular}
\end{center}
\caption{EMCCD specifications}
\label{table:specs}
\end{table}

\section{EXOPLANET SCIENCE WITH A PHOTON COUNTING DETECTOR}\label{sec:science}
\begin{figure}[t!]\centering
 \begin{subfigure}[b]{0.45\textwidth}
  \includegraphics[width=\textwidth]{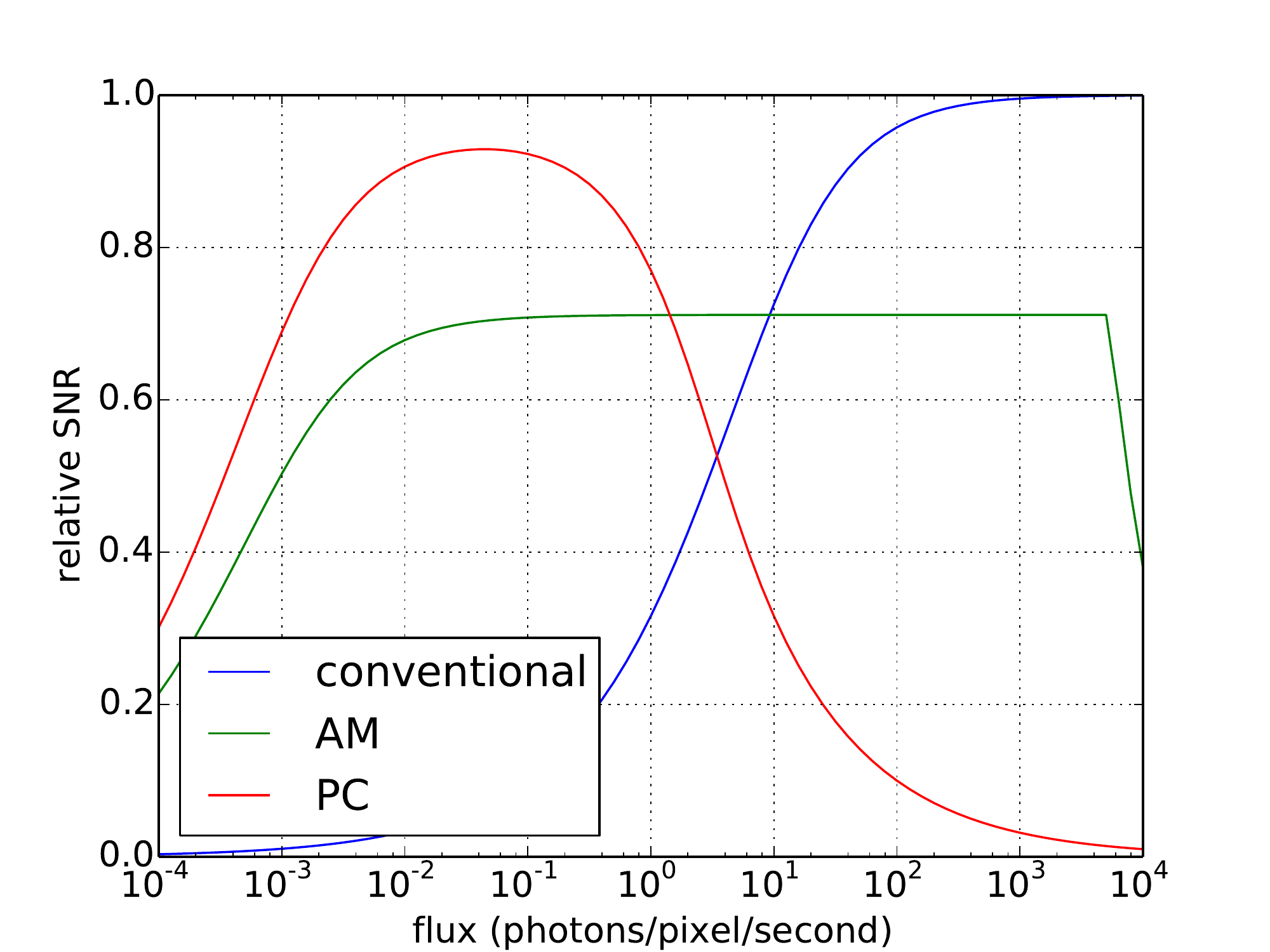}
 \end{subfigure}
 ~
	\begin{subfigure}[b]{0.45\textwidth}
		\includegraphics[width=\textwidth]{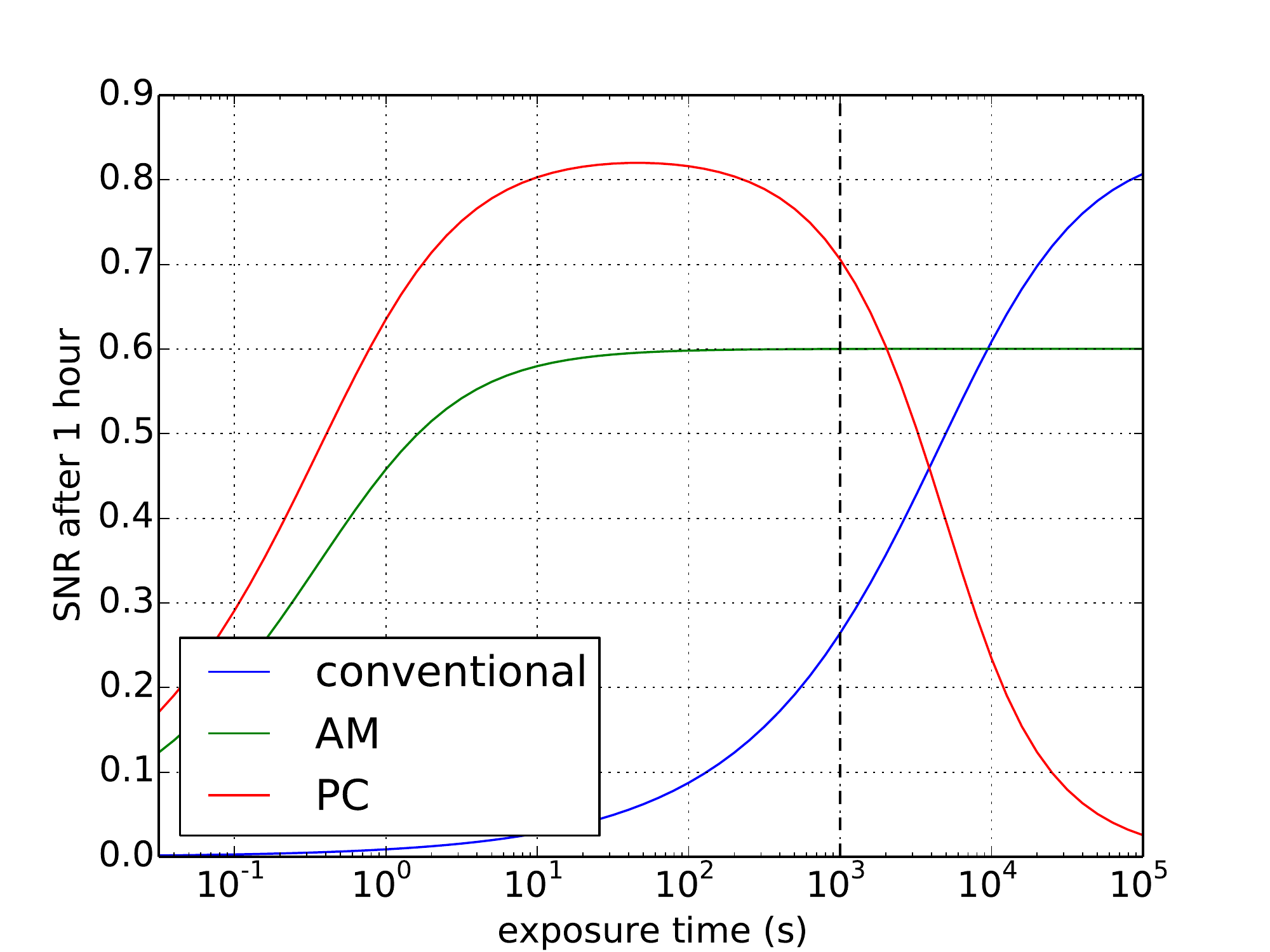}
	\end{subfigure}
 \caption{Models of SNR comparisons in the different operating modes of an EMCCD. \emph{Left:} Comparison of SNR relative to the SNR of a noise-free (Poisson-limited) photon counting device. Modeled is a 536-EM-stage EMCCD with a full well of 800,000 electrons, read noise of 3 electrons, a dark current of 0.0001 electrons/pixel/s and CIC of 0.0007 events/pixel/frame, operated at 1~fps, resembling the device characterized experimentally in \S~\ref{sec:TechApproach}. In analog and PC mode, the EM gain is set to 1000. We use a $5\sigma_{read}$ threshold in PC mode. \emph{Right:} SNR in each spectral channel after one hour of integration at various exposure times on a target of 4.4\,$\cdot$\,10$^{-4}$ photons pix$^{-1}$ s$^{-1}$ (estimate for AFTA observation of a Jupiter-like planet at 5~AU separation from a sun-like star at 10~pc). Coronagraphic starlight suppression is assumed to be $10^{-9}$ and thus yields a background contribution equal to the planet flux. In conventional mode, read noise dominates and thus long integration times are preferred. The vertical line at 1000~s denotes a typical integration time for cosmic ray rejection, but this limit depends on the radiation environment. Analog mode behaves the same (up to saturation), but the impact of read noise is lowered by the EM gain. In PC mode, the optimal exposure time is governed by dark current (plus background light) becoming comparable to the CIC noise floor. The device parameters are the same as in the figure at left.}
\label{fig:snr_fps}
\end{figure}

Photon counting detectors become vitally important in low-flux imaging, when conventional CCDs are limited by read noise, and longer exposures are ruled out by dark or sky background noise. The idea of EMCCDs is to overcome the read noise by multiplying the photoelectrically produced electron before digitization. However, this multiplication step introduces new noise sources, and the detector concept of operations changes considerably compared to a conventional CCD. The randomness of the amplification process causes a spread in the output signal that makes it impossible to clearly distinguish individual few-photon events from one another. This disadvantage of the so-called ``analog mode\cite{Daigle2009}", can be represented as an Excessive Noise Factor (ENF\cite{RobbinsHadwen2003}) that reaches a value of 2 for high gains and effectively reduces the signal-to-noise ratio (SNR) by a factor of $\sqrt 2$. In order to overcome this limitation, EMCCDs can be operated in Photon Counting (PC) mode. Here, the frame rate is raised so that the arrival of two photons in the same frame and pixel is unlikely for the expected fluxes. Then, a threshold is introduced to transform pixel values into one bit of information: noise only or photon(s) arrived. Data acquisition in this mode entails the rapid accumulation and co-addition of a sequence of binary frames. The threshold is canonically set to five times the standard deviation (5$\sigma$) of the read noise, though we intend to investigate this parameter more carefully in the future. The threshold avoids significant false counts, while the large majority of electron events can be amplified above it with realistic gains. At all gain settings the probability distribution for the amplification of a single electron extends to zero additional output electrons. Thus, there is a non-vanishing chance of a photon signal staying below the 5$\sigma$ threshold. The electron multiplication (EM) gain must be set to more than 50 times the read noise in order to count 90\% of the incoming photons~\cite{Daigle2009}. This follows directly from the statistics of the amplification process.

While the technique of thresholding eliminates the ENF and enables photon counting, SNR deteriorates at high fluxes due to coincidence losses.
Additionally, the problem of clock-induced charge (CIC) emerges from the noise floor. CIC are spurious electrons generated in the readout process, and thus proportional to frame rate for fixed integration time. In conventional CCDs the CIC is present but negligible compared to read noise and therefore eludes detection. The high-voltage clocking in the EM register as well as the high frame rate make CIC the limiting factor for EMCCD sensitivity at low signal, whether at high framing~\cite{MacKay2012,Daigle2010} or at long exposures~\cite{Daigle2010}.

The dynamic range of a conventional CCD is given by the ratio of full well depth to readout noise. In analog, the EMCCD is still limited by read noise and saturation, but now the read noise does not get amplified while the signal does, and it is the full well in the EM multiplication stages that limits the output signal, long before the image region pixels saturate. Therefore, both the effective read noise and the effective full well are the respective conventional values divided by the EM gain. Their ratio, the dynamic range in AM, is the same as in a conventional CCD. The EM gain shifts the sensitive range of the EMCCD towards lower fluxes without impacting dynamic range, until dark noise or CIC (which is amplified in the EM register) becomes limiting.

In a photon counting EMCCD, both the upper and lower sensitivity limits are set by new constraints. The CIC level determines the lowest fluxes that can be measured (typically several orders of magnitude below the conventional read noise). At the high-flux end, an EMCCD in photon counting mode loses sensitivity due to its thresholding strategy which does not allow the discrimination between multiple photon events. Poisson statistics allow us to calculate the fraction of lost photons by coincidence $f_{c}$ as a function of mean arrival rate $\mu$ in photons per pixel per frame:
\begin{equation}
f_{c} = \frac{\sum_{i=2}^{\infty}(i-1)P_\mu (i)}{\sum_{i=0}^{\infty}iP_\mu(i)} = 1-\frac{1-e^{-\mu}}{\mu}
\end{equation}
For example, to restrict coincidence losses to less than 5\% of the signal, the EMCCD has to be operated at an expected signal rate of less than 0.1 photons per pixel per frame. Assuming a CIC level of 0.0007 photons per pixel per frame, the dynamic range evaluates to 143:1 (5.4 mag).

Changing the exposure time allows the observer to shift the dynamic range about some offset point. The CIC contribution to an observation of fixed duration is proportional to frame rate, but higher frame rates allow proportionally higher fluxes to be detected at a fixed level of coincidence loss. Ultimately, the range over which an EMCCD can be used in photon-counting mode depends on the maximum readout speed for brighter applications and detector dark noise for low-flux imaging with long integration times.

A single EMCCD detector can be switched between all three operation modes (conventional, AM, PC mode), since the threshold in PC mode is software-defined and conventional mode differs from the other two only by the amplitude of the high-voltage clock in the gain register.

The SNR of a conventional CCD is given by
\begin{equation}
SNR_{conv} = \frac{S}{\sqrt{S+D+\sigma_{r}^2}}
\end{equation}
with the expectation value of the signal electrons $S$, dark current electrons $D$ and read noise $\sigma_{r}$. In AM, the SNR changes to
\begin{equation}
SNR_{AM} = \frac{S}{\sqrt{F^2S+D+\frac{\sigma_{r}^2}{g^2}}}
\end{equation}
with electron multiplying gain $g$ and ENF $F^2$, where
\begin{equation}
F^2 = 2(g-1)g^{-\frac{N_s-1}{N_s}}+\frac{1}{g}
\end{equation}
with the number of EM stages $N_s$. For a detailed derivation see~\cite{RobbinsHadwen2003}.

The SNR in PC mode has been estimated as follows. Read noise only impacts SNR via false counts that occur when a fluctuation causes a dark pixel to exceed the threshold. These false counts can be suppressed effectively by setting the threshold to a value high enough above the mean read noise, typically $5\sigma_{r}$~\cite{Daigle2009}. Then, the expected read noise contribution is below $10^{-6}$ photons/pixel/frame and can be neglected (since it is much smaller than the contribution of clock-induced charges described below). The threshold also causes two new kinds of losses: photons that fail to get multiplied above the threshold (threshold loss $\eta_t$) and coincidentally arriving photons that result in only one count (coincidence loss $\eta_c$). Additionally, the noise contribution of clock-induced charge (CIC) now becomes non-negligible because it is multiplied in the same way that a photoelectrically generated electron is amplified. 

Within a frame, the events are binary and can not be described by a Poisson distribution. In spite of this, the counts are distributed among the co-added binary frames according to Poisson statistics, and thus the previously used statistical framework can be applied to the PC mode as well. Since dark current and CIC events are determined from binary frames in PC mode, no threshold losses have to be incorporated because only events that exceed the threshold affect the signal to noise calculation. Coincidence losses between dark, CIC and signal are a higher-order effect and have been neglected in this discussion.
\begin{equation}
SNR_{PC} = \frac{\eta_t\eta_cS}{\sqrt{\eta_t\eta_cS+D+CIC}}
\end{equation}
with coincidence loss
\begin{equation}
\eta_c = \frac{1-\mathrm{e}^{-S}}{S}
\end{equation}
The threshold loss $\eta_t$ is calculated from the photon arrival and EM amplification probability distributions, as a sum truncated at the threshold value~\cite{Daigle2009}.

At left in Figure~\ref{fig:snr_fps}, SNRs for the same EMCCD operated in conventional, analog and PC modes are compared. In conventional mode, SNR decays quickly towards low fluxes due to the larger relative contribution of read noise. Electron multiplication overcomes the read noise, at which point it becomes apparent how analog mode now suffers from the ENF. PC mode removes the ENF by introducing the threshold and obviating the need for a statistical lookback. However, PC mode then is penalized by a reduction in dynamic range due to the coincidence losses. Each of the modes has a corresponding regime where its sensitivity is the highest, and it is therefore important to identify the regime corresponding to a given set of observations. The PC mode is oftentimes the best choice unless the flux rate exceeds the fastest sampling rate (i.e., the sampling rate can not keep up with the photon arrival rate) or the observations are in need of high dynamic range. 

The superior sensitivity of photon counting EMCCD detectors is crucial in the field of exoplanet direct imaging, as the expected fluxes are extremely low. For example, an integral field spectrograph instrument with a bandpass from 400~nm to 1000~nm and a spectral resolution of $R\sim70$ (as currently planned for the AFTA coronagraph\cite{Shaklan2013}) will receive a flux of only 4.4\,$\cdot$\,10$^{-4}$ photons pixel$^{-1}$ s$^{-1}$ from a planet orbiting a 6000~K, 700000~km (radius) blackbody at a distance of 10~pc. This assumes two pixels per resolution element, a planet-to-star contrast of $10^{-9}$ (i.e. a Jupiter-like planet at 5~AU separation) and the AFTA telescope dimensions and throughput~\cite{Shaklan2013}. Coronagraphic starlight suppression is assumed to be $10^{-9}$ and thus yields a background contribution equal to the planet flux. From the SNR estimates for this instrument shown at right in Figure~\ref{fig:snr_fps}, the integration times necessary to achieve a SNR of 5 in each spectral channel for the three operation modes are calculated and compared with each other. At the optimal exposure times for each operating modes (50~s in PC, 65~s in analog mode, and a maximum reasonable 1000~s in conventional mode), the necessary integration times are 36 hours (PC), 69 hours (analog) and 15 days (conventional). For the 16~m ATLAST~\cite{Postman2010} observing a target with planet-to-star contrast of $10^{-10}$ for an Earth-like planet (adjusting the optimal exposure times), the integration times change to 76 days (PC), 140 days (analog) and 350 days (conventional) to reach an SNR of 58 for biomarker detection~\cite{Heap2008}. This assumes improved coronagraphic starlight suppression of $10^{-10}$ and matching of AFTA's throughput performance. This clearly shows the necessity of photon counting for an exoplanet-characterizing space mission.

\section{DETECTOR CHARACTERIZATION}\label{sec:TechApproach}
\begin{table}
\begin{center}
 \begin{tabular}{| c | c | c | c |}
 \hline
 Parameter & N\"uv\"u Reported Values & Measured Values & Measured Values \\
 & & (Conventional Mode) & (EM, PC Mode) \\ \hline \hline
 Conventional Gain & N/A & 4.8 e$^{-}$/ADU & N/A \\ \hline
 Dark Current & 0.0002 e$^{-}$/pixel/s & 0.000084 e$^{-}$/pixel/s & 0.000072 e$^{-}$/pixel/s \\ 
 (T=-85$^{o}$C) & & & \\ \hline
 CIC & \textless 0.001 e$^{-}$/pixel/frame & N/A & 0.00076 e$^{-}$/pixel/frame\\ 
 (T=-85$^{o}$C, EM Gain=1000) & & & \\\hline
 \end{tabular}
\end{center}
\caption{Reported Values vs. Measurements}
\label{table:compare}
\end{table}
While the performance of the e2v CCD97 EMCCD on its own has been evaluated \cite{Wen2006}, the performance of the detector within the N\"uv\"u camera housing and controller has not been independently investigated. Especially given the significant improvements N\"uv\"u has claimed in areas such as dark current and CIC, an independent analysis of the performance will only aid in raising confidence and lowering risk for use of such cameras in future astronomy applications. To that end, we have begun extensive characterization, the progress of which is summarized here. 

\begin{SCfigure}\centering
 \begin{subfigure}[b]{0.3\textwidth}
  \includegraphics[width=\textwidth]{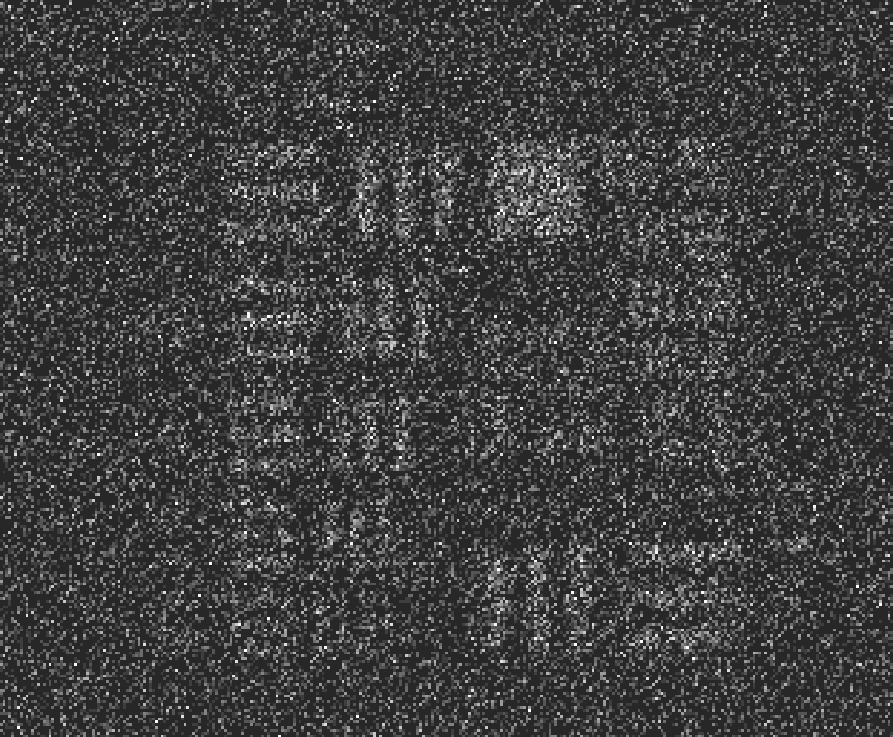}
 \end{subfigure}
 ~
	\begin{subfigure}[b]{0.3\textwidth}
		\includegraphics[width=\textwidth]{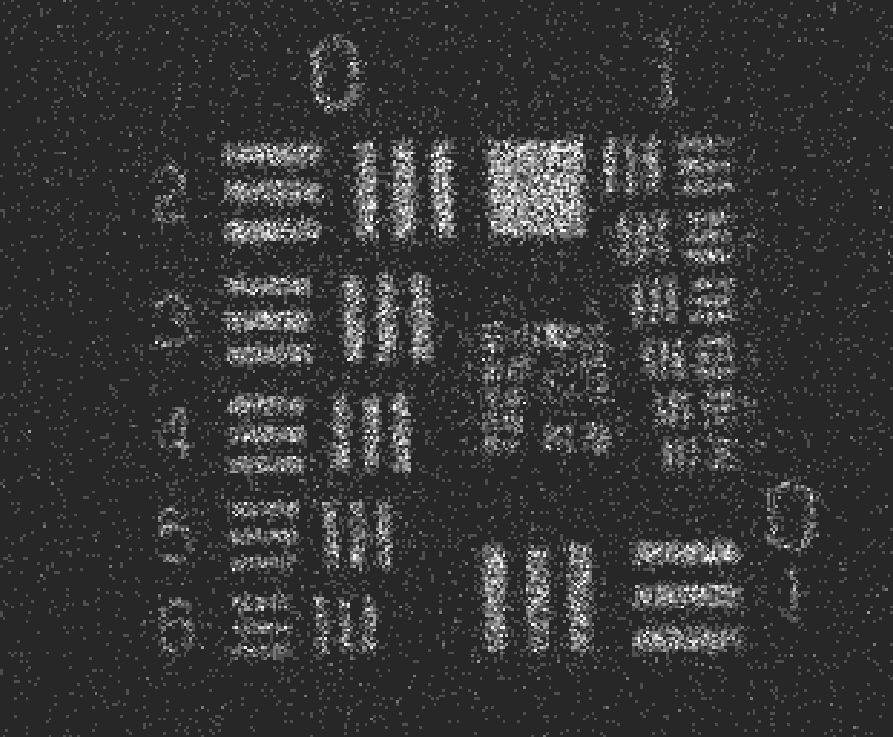}
	\end{subfigure}
 \caption{A standard USAF target imaged by the N\"uv\"u HN\"u 512. The target's illumination is very low and the same in both panels. Both images are also scaled linearly. \emph{Left: }Conventional mode operation, one ten-second frame. \emph{Right: }Photon counting mode operation, twenty 0.5-second frames, for a total integration time also of ten seconds, but a significant improvement in SNR.}
 \label{fig:conv_vs_pc}
\end{SCfigure}

First, we show the superior performance of photon counting over conventional operation in Figure~\ref{fig:conv_vs_pc}. This qualitative visualization of the quantitative explanation from \S~\ref{sec:science} shows the significant gain in SNR when using photon counting to image a faint target, and the performance specifically of the e2v CCD97 in the HN\"u 512. 

For quantitative analysis, we began in conventional mode, then transitioned to EM mode. Some of the key values are given in Table~\ref{table:compare}, and our methods for obtaining those values are described below. 

\subsection{Conventional Mode}
We begin with the conventional mode operation of the CCD, calculating the standard parameters of conventional gain and dark current. For all conventional mode operation, we clocked the detector at a horizontal frequency of 3.33 MHz and a vertical frequency of 1MHz. 

\subsubsection{Photon Transfer Curve}
We use the standard method of taking the photon transfer curve to measure the gain. The plot in the left panel of Figure~\ref{fig:ptc_dark} shows this curve, with the variance of the difference of two flat fields (y-axis) plotted as a function of the mean value of those two flat fields. The slope of the linear realm of this plot is the inverse of the conventional gain, which we find to be 4.8 electrons/ADU. 

\subsubsection{Dark Current}\label{sec:dark}
The N\"uv\"u camera utilizes Inverted Mode Operation (IMO) to significantly reduce the dark current as opposed to the standard Non-Inverted Mode Operation (NIMO). In IMO, the vertical transfer clock voltages are set to be more negative than the substrate to suppress interface states at the SiO$_{2}$ - Si interface by attracting holes that ``mop up"\cite{Tulloch2010} the electrons which would otherwise contribute to dark current. We illustrate this mechanism in Figure~\ref{fig:cic_mechanism}. The result is a significantly lower dark current, but a somewhat higher CIC, thought to be due to impact ionization as the CCD is moved in and out of inversion. N\"uv\"u uses other methods to combat the CIC (see \S\ref{sec:CIC}), and thus produces an overall improvement by utilizing IMO instead of NIMO. 

The HN\"u 512 does have minor light leaks, to the point that simply closing the shutter is not sufficient for the long (multi-hour) exposure times required to measure any significant number of dark current electrons. Thus, we operated the camera under a partially open dark box, and with several layers of black cloth covering the shutter side of the camera.

\begin{figure}\centering
 \begin{subfigure}[b]{0.45\textwidth}
  \includegraphics[width=\textwidth]{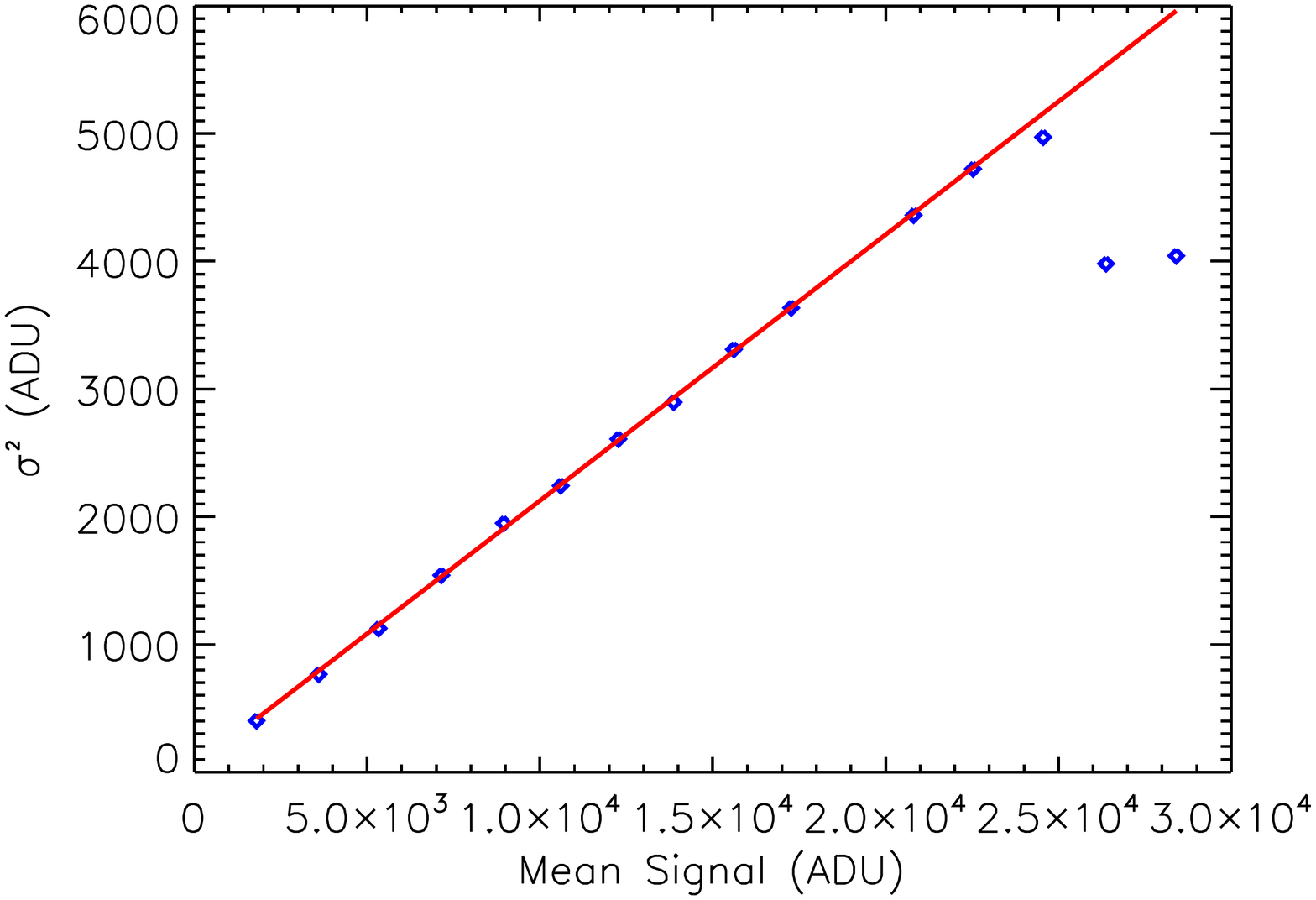}
 \end{subfigure}
 ~
	\begin{subfigure}[b]{0.45\textwidth}
		\includegraphics[width=\textwidth]{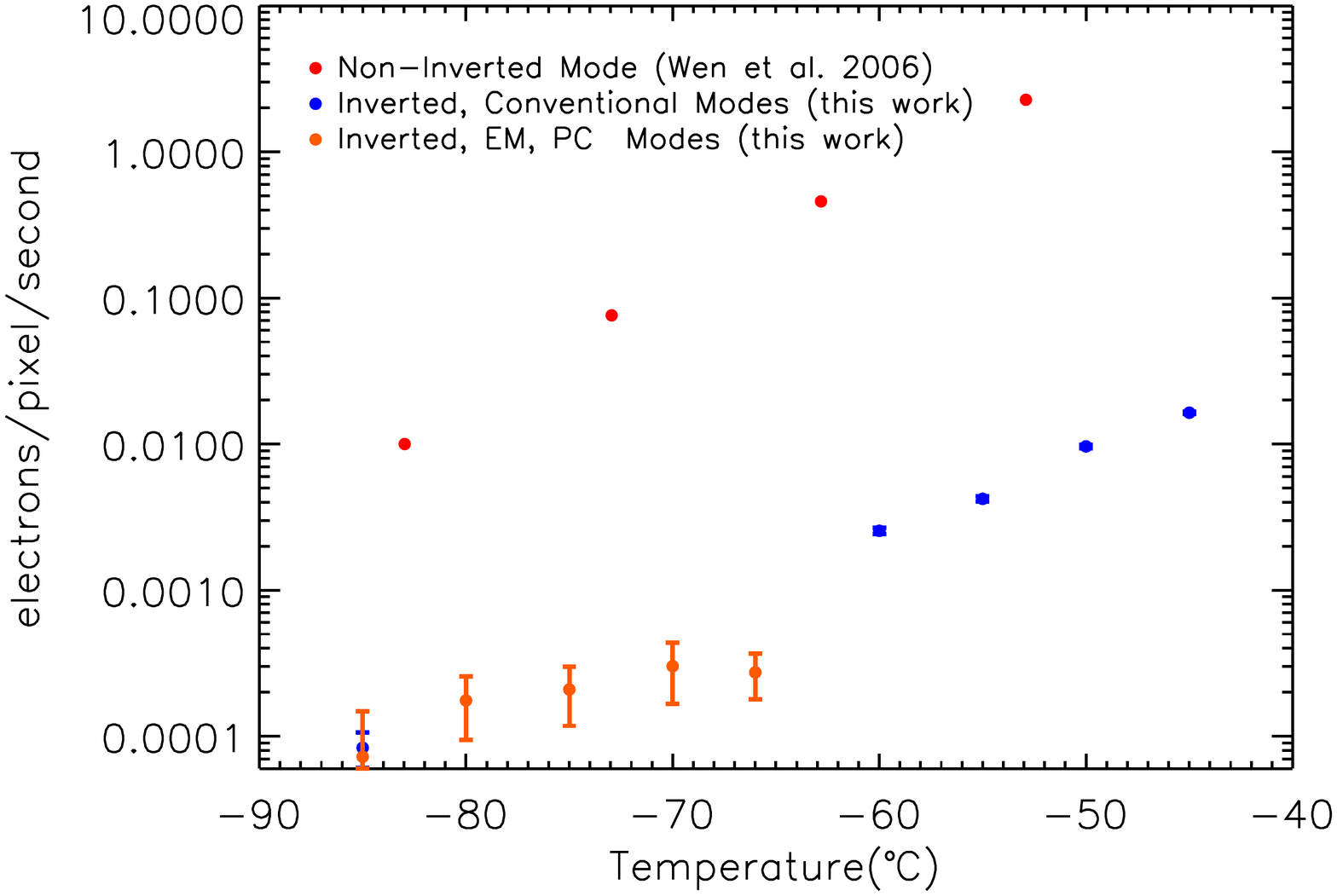}
	\end{subfigure}
 \caption{Detector characterization. \emph{Left:} Photon transfer curve for conventional gain calculation. Blue points are data taken from flat images at increasing exposure times, and the red line is the fit to the linear data, the slope of which is the inverse gain. \emph{Right:} Dark current as a function of temperature. For comparison, the dark current from the same detector model operated in NIMO is plotted in red. Dark current measured in conventional mode is in blue, and in photon counting mode in orange.}
 \label{fig:ptc_dark}
\end{figure}

We confirm the very low dark current measurement reported in Table~\ref{table:compare}, and show the dark current as a function of temperature in the right panel of Figure~\ref{fig:ptc_dark}. For comparison, we plot dark current measurements of another CCD97 operated in NIMO\cite{Wen2006}, which were two orders of magnitude higher. 

\subsection{EM Mode}
When operating in Electron Multiplying (EM) mode, all electrons transferred from the imaging area of the detector pass through the gain register. This includes electrons not generated photoelectrically by an incident photon, a significant noise contribution described in \S\ref{sec:CIC}. Except where otherwise noted, we clocked the detector at a horizontal frequency of 20 MHz and a vertical frequency of 3.33 MHz.

\subsubsection{Dark Current} 
The dark current generated in the image area should have no dependence on the operating mode of the camera; it should remain constant throughout conventional and electron-multiplying modes. However, the dark current generated in the electron multiplying gain register may be dependent on the gain.  Measuring the dark current is less time-intensive in PC mode than in conventional mode, as signal can be reliably measured in much shorter exposures, thanks to the gain register. The orange points in the plot of Figure~\ref{fig:ptc_dark} represent these measurements at several temperatures, using an EM gain of 1000. The dark current at temperatures warmer than -66$^{o}$C cannot be measured in this way, because the N\"uv\"u software prevents any EM gain at the warmer temperatures. This was a conservative precaution to avoid saturation and damage for the EM register, and will be relaxed in later releases of the software. 
We measure dark current at -85$^{o}$C at approximately 8$\times$10$^{-5}$ electrons/pixel/second, very similar to the value found in conventional mode. 
\begin{SCfigure}\centering
 \includegraphics[scale=0.3]{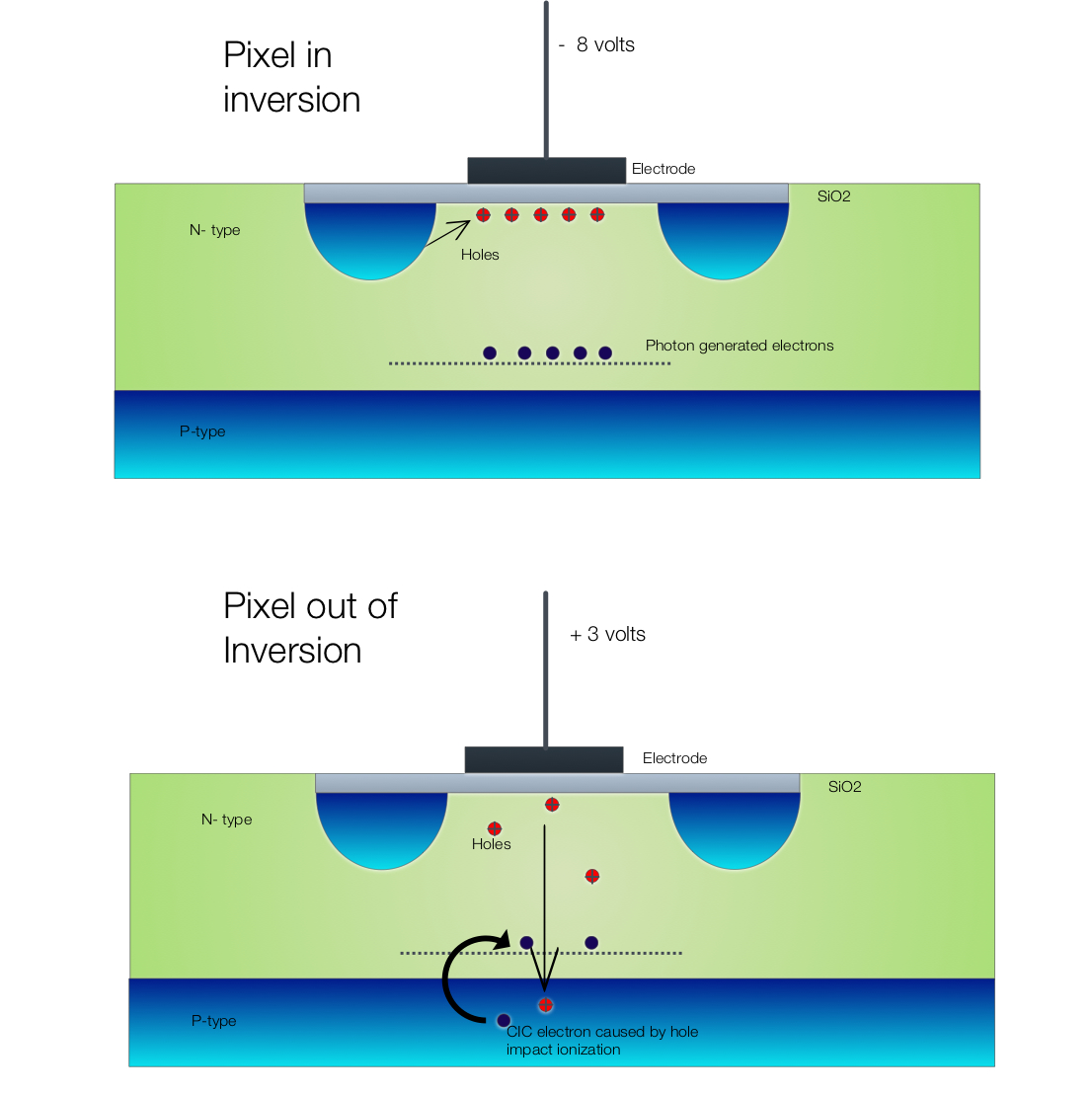}
 \caption{An illustration of the influence the operation mode (NIMO/IMO) has on holes. In IMO, holes are attracted to the gate and they ``mop up" many of the electrons that would otherwise contribute to dark current. Movement from the inverted to the non-inverted state in the process of clocking the detector is also the mechanism responsible for forming surface-layer CIC, which is produced as the voltage changes.}
 \label{fig:cic_mechanism}
\end{SCfigure}

\subsubsection{Clock-Induced Charge}\label{sec:CIC}

All CCDs exhibit clock-induced charge. The mechanism for generating surface-layer CIC due to clocking between IMO and NIMO is illustrated in Figure~\ref{fig:cic_mechanism}. As the device is read out and the clocks come out of inversion, the holes, under high electric field, are driven into the silicon where they can, by impact ionization, create secondary electrons - the clock-induced charge. As alluded to earlier, in conventional mode, with read noise of the order 3 e$^{-}$, these electrons are hidden in the noise. However, in EMCCDs with gain greater than 1000 these spurious events can be observed and are indistinguishable from true photoelectrically-induced electron events. CIC is also observed in the serial EM register. These events tend to be generated, on average, half-way down the 536-element EM gain stage register and thus are of smaller amplitude than image-area-generated CIC. 

Operating completely in NIMO significantly lowers CIC, but even more significantly raises the dark current, as described in \S~\ref{sec:dark}. Instead, the transition between IMO and NIMO can be adjusted with waveform shapes that minimize the CIC, as the pulse rise time and the sharpness with which it is changed are primary influences on the CIC generation. Advanced waveform shaping and control is implemented by the N\"uv\"u CCD Controller for Counting Photons (CCCP\cite{Daigle2009,Daigle2010,Daigle2012}) and is primarily responsible for their extremely low combined dark current and CIC floor, as compared to the operation of the CCD97 EMCCD alone. 

\begin{figure}\centering
 \begin{subfigure}[b]{0.45\textwidth}
  \includegraphics[width=\textwidth]{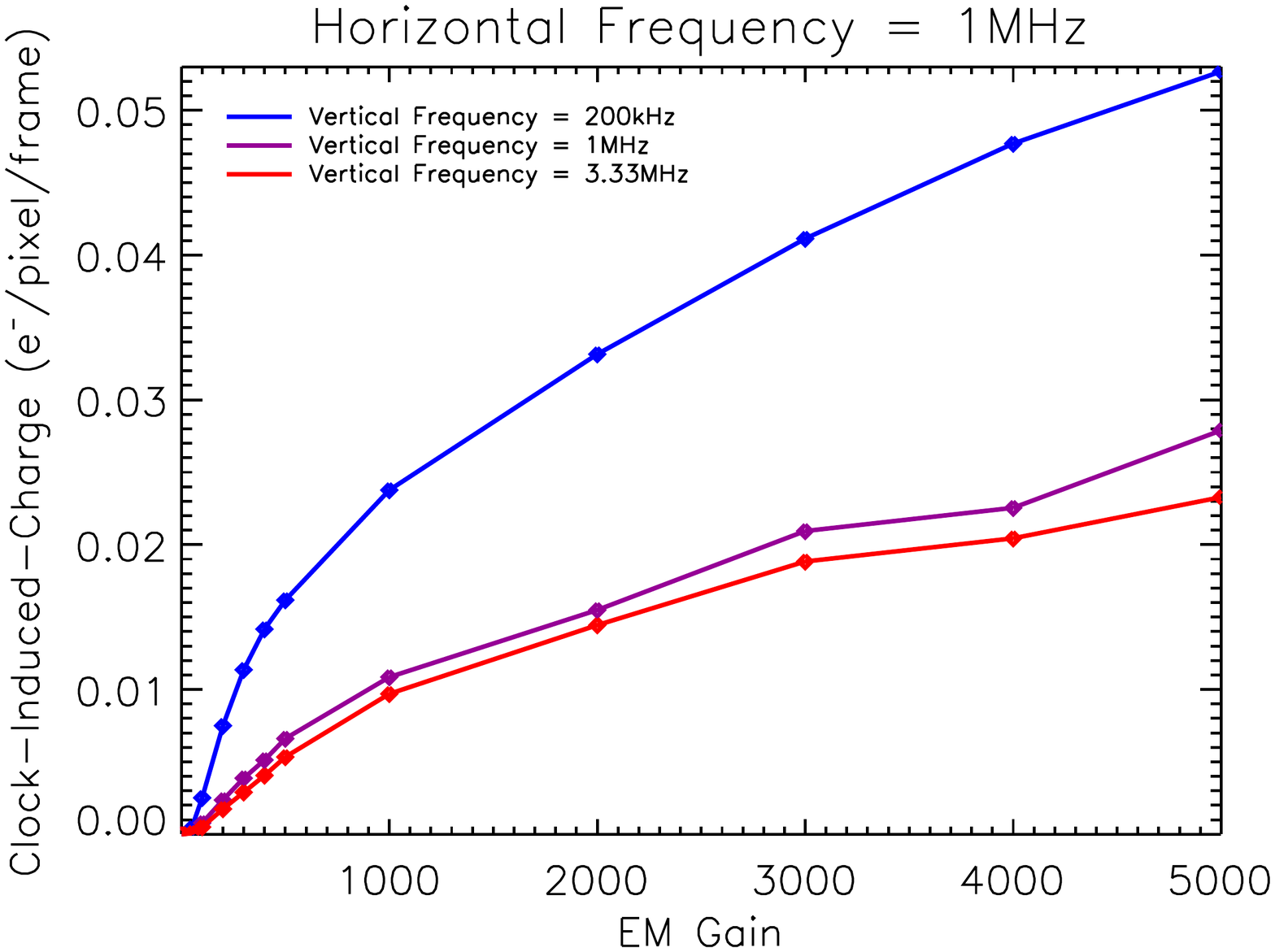}
 \end{subfigure}
	\begin{subfigure}[b]{0.45\textwidth}
		\includegraphics[width=\textwidth]{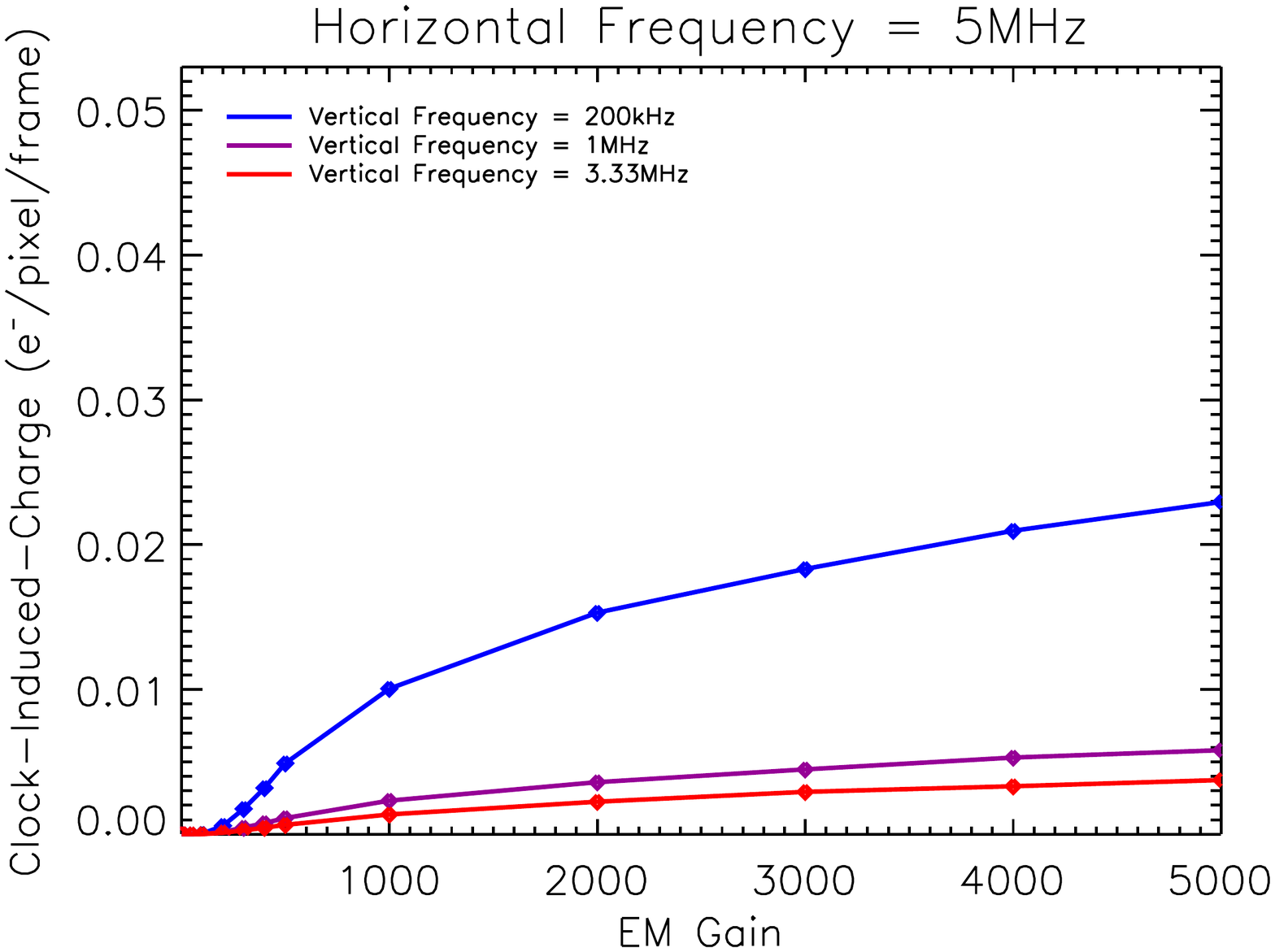}
	\end{subfigure}
	\begin{subfigure}[b]{0.45\textwidth}
		\includegraphics[width=\textwidth]{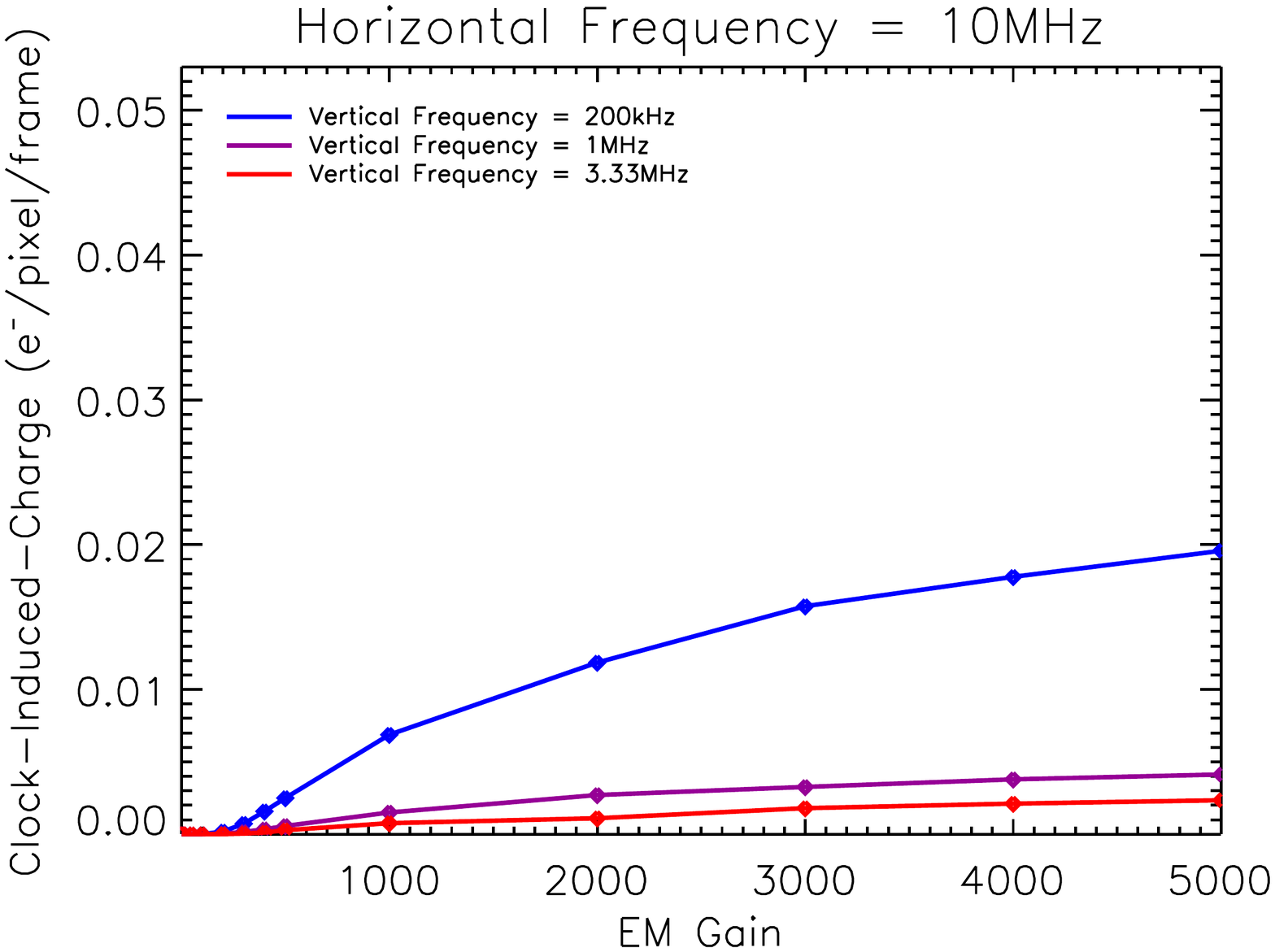}
	\end{subfigure}
	\begin{subfigure}[b]{0.45\textwidth}
		\includegraphics[width=\textwidth]{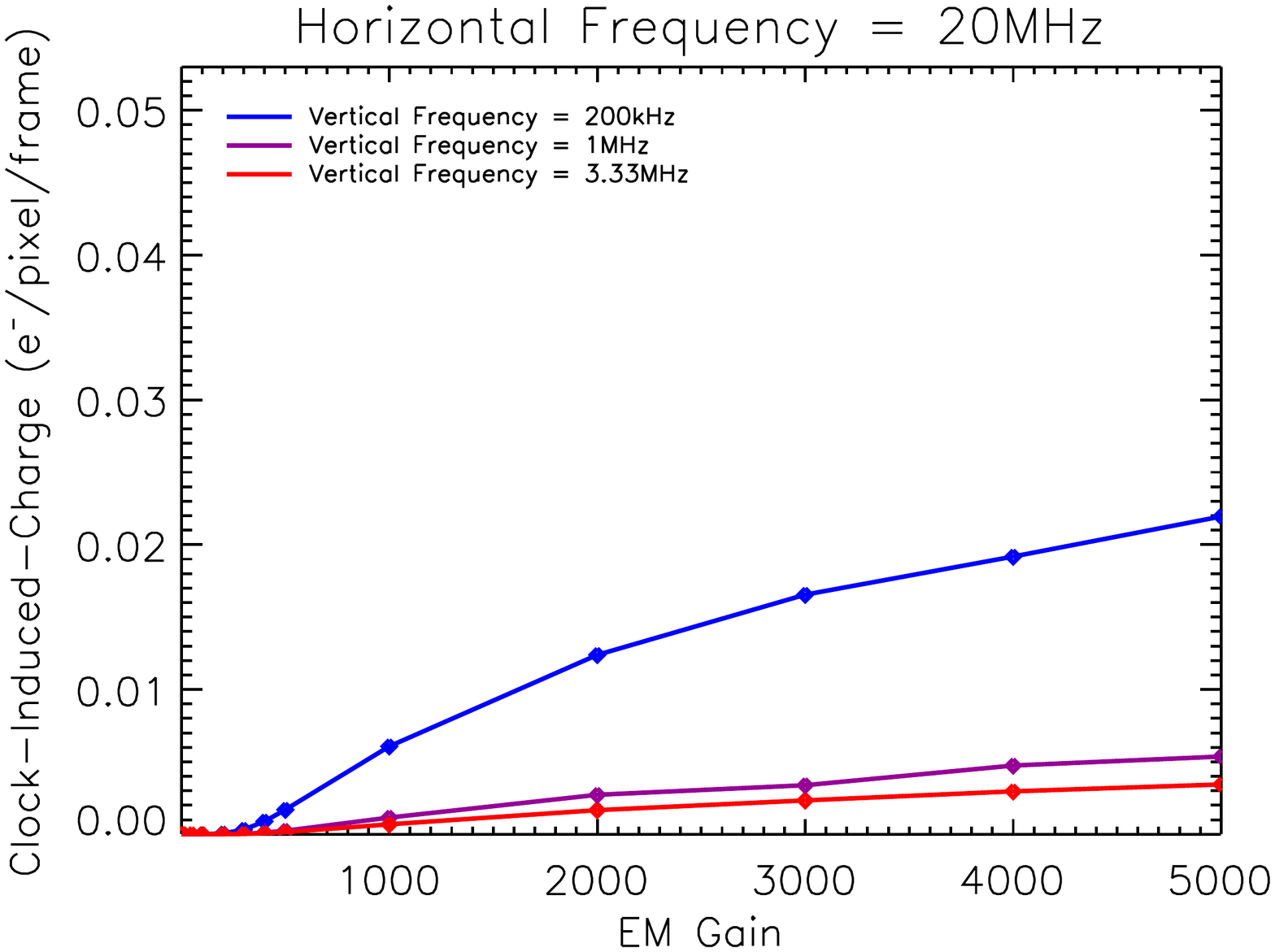}
	\end{subfigure}
 \caption{CIC in electrons/pixel/frame as a function of EM gain at various horizontal and vertical clocking frequencies. The slowest clocking allows 4 frames per second, while the fastest clocking allows 67 frames per second. Faster clocking mitigates CIC, to similar degrees in either dimension.}
 \label{fig:cic_rates}
\end{figure}

When operating at -85$^{o}$C and an EM gain of 1000, we measure a CIC of 0.0007 electrons/pixel/frame, which is below, but close to, the N\"uv\"u-reported value of 0.001 electrons/pixel/frame (see Table~\ref{table:compare}). We measured CIC at a range of EM gains and horizontal (or serial) and vertical (or parallel) clocking speeds, demonstrating the known behavior that CIC increases with increasing gain and decreases with increasing clock speeds. These data are plotted in Figure~\ref{fig:cic_rates}. 

\section{APPLYING SUB-BANDGAP ILLUMINATION TO AN (EM)CCD}\label{sec:sub-bandgap}
Our hypothesis is that sub-bandgap illumination will dramatically decrease the decay time of trapped electrons that would otherwise be released upon application of significant voltage (e.g., during clocking) and indistinguishable from charges photoelectrically generated from a target source photon. By ``sub-bandgap illumination", we mean infrared photons with energy values below that of the silicon bandgap, and thus unable to photoelectrically generate an electron. Visible background light has been shown to improve the CTE of radiation-damaged EMCCD pixels\cite{Michaelis2013b}, but that method injected background light that would be undesirable for astrophysics. We introduce a sub-bandgap source in the field of view of the detector and therefore do not create an external background in the frames.

We note that the sub-bandgap illumination will not impact the EM gain register, as it is shielded with aluminum from any such photons by design. This shielding is not a fundamental necessity, however, and may be deemed undesirable in a future custom-designed package. This has two implications: first, that any work performed here applies to a conventional CCD (and conventional CCD operation) just as it does to an EMCCD (and EM, PC-mode operation), and second, any CIC actually due to traps or other factors within the gain register will not be affected. With those caveats in mind, we discuss the progress of two experiments meant to demonstrate an interaction of infrared photons with traps and the electrons that leak from them. 

For both tests, the camera, optical light source, and infrared light source were all enclosed within a dark box, which was then covered with dark cloth, on a breadboard in a darkened room. For both experiments, we clocked the detector at a horizontal frequency of 20 MHz and a vertical frequency of 3.33 MHz.

\subsection{First Attempts at Direct CIC Mitigation}\label{sec:cic_mitigate}

A very preliminary test for the response of CIC to sub-bandgap illumination is to measure the CIC as a function of a source's temperature, beginning with no illumination, increasing the temperature to the point of producing thermal photons below the detector bandgap energy, then finally reaching detectable optical signal at high temperatures. We chose a standard lightbulb to perform this experiment, and used increasing voltage which represented a proxy for temperature. 

We connected the same projector bulb used to generate the images in Figure~\ref{fig:conv_vs_pc} to a Variac transformer, which allowed a fine-tuning of the light bulb's voltage, and therefore its temperature and brightness. The Variac was outside of the dark enclosure for efficiency in adjustment. Measurements were taken at intervals of 0.1 V, measured with a voltmeter on the Variac from zero illumination through the point of optical signal detection. 

\begin{figure}\centering
 \begin{subfigure}[b]{0.45\textwidth}
  \includegraphics[width=\textwidth]{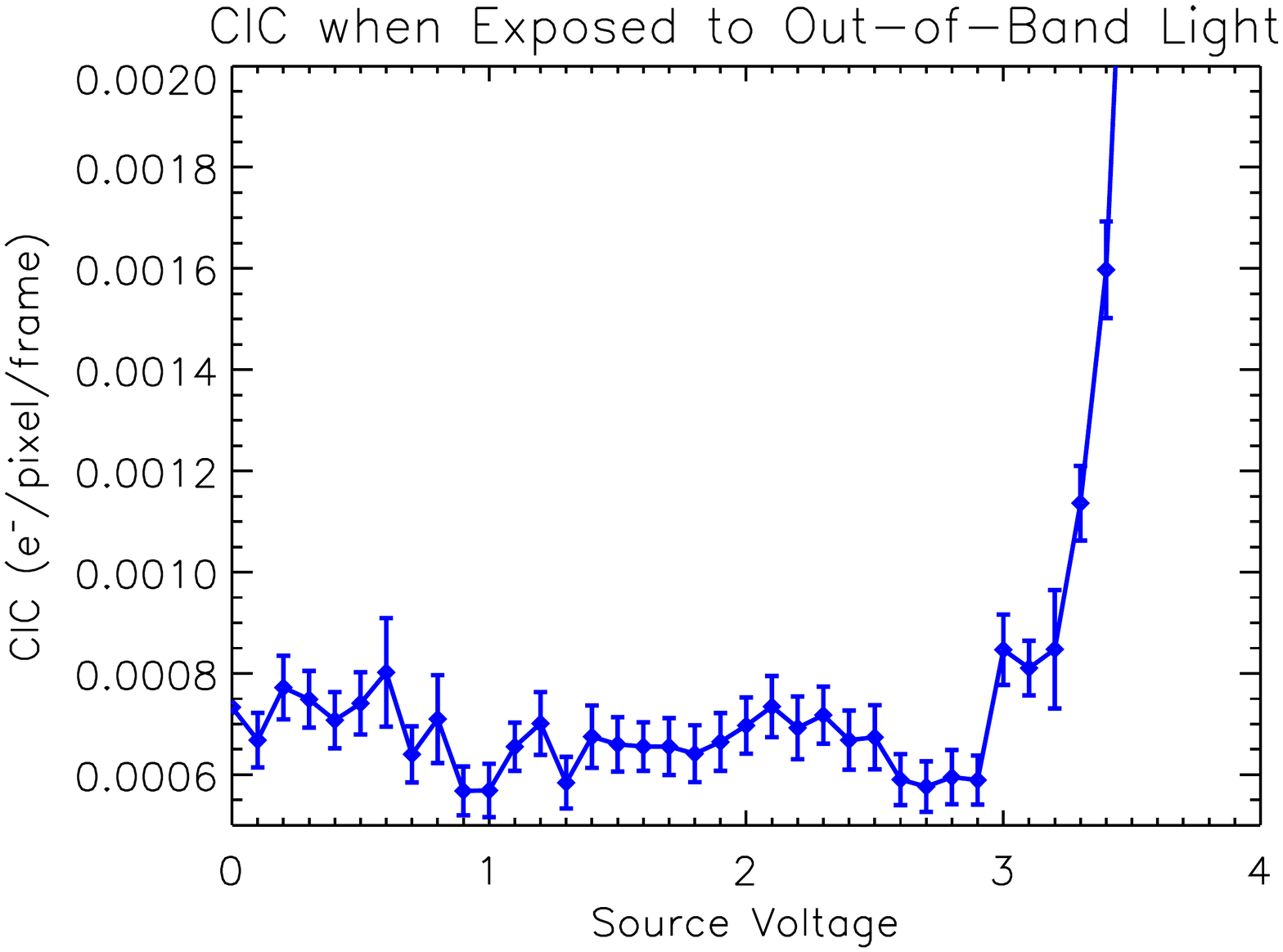}
 \end{subfigure}
 ~
	\begin{subfigure}[b]{0.45\textwidth}
		\includegraphics[width=\textwidth]{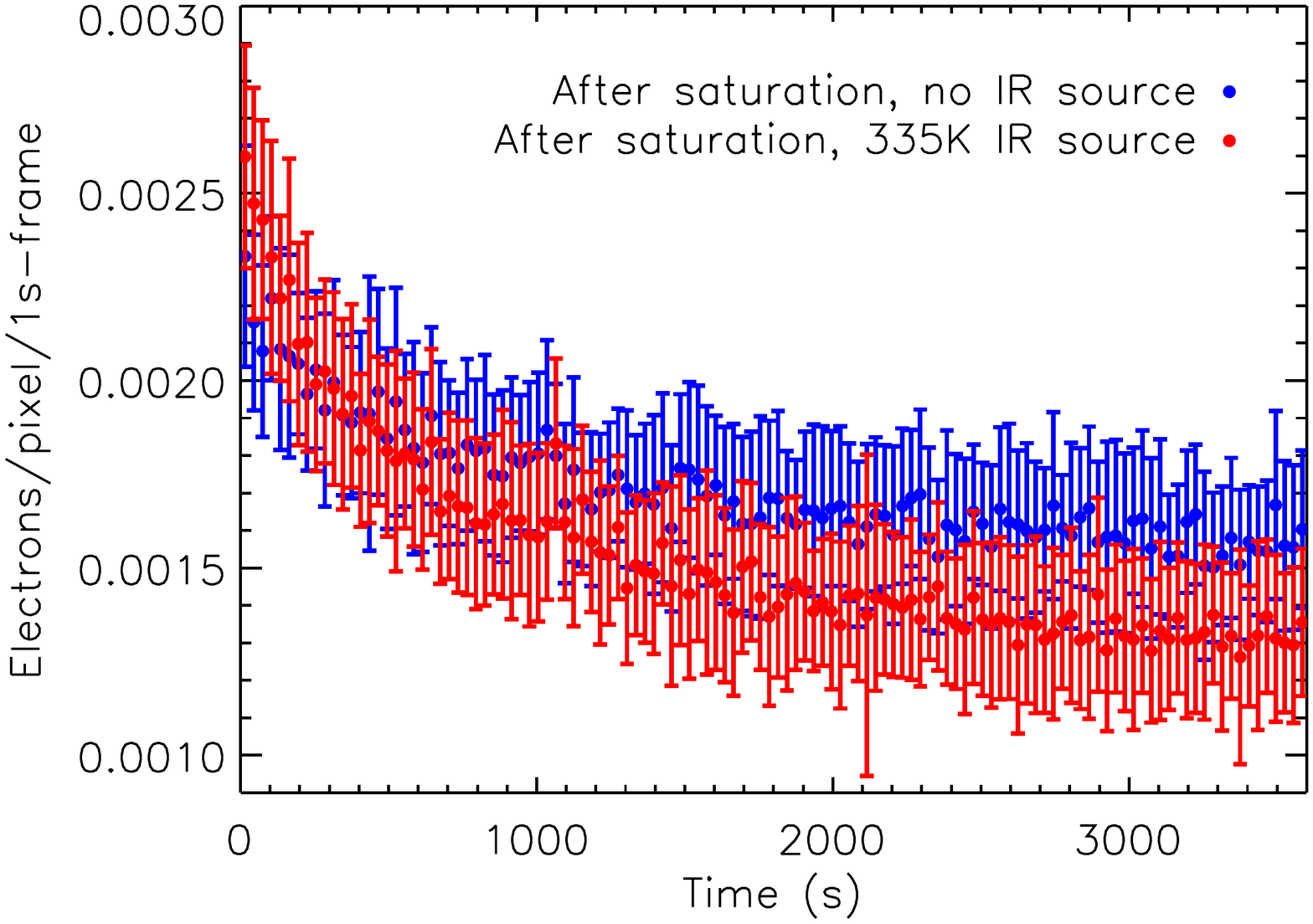}
	\end{subfigure}
 \caption{Results from sub-bandgap illumination experiments. \emph{Left:} CIC as a function of voltage sent to IR source. There is no statistically significant drop in CIC before the spike indicating the presence of optical photons. \emph{Right:} Persistence signal as a function of time after fifteen minutes of saturation. One-second frames are binned to thirty second. When a 335K source is present (red points), the signal is higher initially, as electrons are being quickly released, and then it drops more steeply than when no IR source is present (blue points).}
 \label{fig:sub-bandgap}
\end{figure}

The results of this experiment are plotted in the left panel of Figure~\ref{fig:sub-bandgap}. From the zero-voltage point to the spike where optical photons are being detected at approximately 3.0V, no statistically significant reduction in CIC appears. This could be for two reasons: 
\begin{enumerate}
\item The rate of infrared photons was too low to make a significant difference in stimulating the traps. 
\item The wavelength of the infrared photons needs to be more finely tuned to stimulate the traps, perhaps specifically to much shorter wavelengths. 
\end{enumerate}
To address option 1, we simply need to find a stronger source, the options for which are numerous. To address option 2, we will make similar experiments with LEDs emitting at narrow infrared wavelengths. For now, we consider this an inconclusive result.

\subsection{Measuring Interaction of Sub-Bandgap Illumination through Persistence Decay}

For the second test, we looked at the well-known phenomenon of persistence, or Residual Bulk Image (RBI), which occurs after a bright source has filled the wells approximately $\frac{2}{3}$ full and results in residual images in later reads, even when the source is removed. The residual image slowly fades away as electrons are released. The timescale for this release increases as detector temperature decreases\cite{Rest2002}, and for operation at -85$^{o}$C, this may be hours. For this test, we operated at -66$^{o}$C, warmer for the shorter, more practical decay timescale, but cold enough to make use of the gain register (recall that the N\"uv\"u software does not allow EM gain at temperatures above -65$^{o}$C as a precaution). 

Our hypothesis in this case was that sub-bandgap illumination applied after saturation would speed up the decay of the electrons causing persistence by giving them enough energy to escape from the traps. Persistence would be worse immediately after saturation, as more electrons are released initially, but then would fade more quickly to approach the original zero point.

To saturate the detector, we again used the projector and Variac system, allowing us to turn on the light to saturation levels and turn it off again without opening the dark enclosure. For the heat source, however, we changed from the first experiment to using a soldering iron directly in front of the camera, but out of the light path of the projector. The benefits of using the soldering iron over the bulb included being able to directly probe the source temperature (by attaching thermocouple wires directly to the tip) and getting the source as close to the detector as possible. 

The experimental procedure began with saturating the detector in conventional mode for fifteen minutes of integration. We then switched to EM/PC mode and took consecutive one-second frames for one hour. In the initial round, the soldering iron was off. We then repeated the procedure, but turned the soldering iron on to approximately 335K. 

The results of this experiment, binned to 30 seconds, are in the right panel of Figure~\ref{fig:sub-bandgap}. From that plot, we found agreement with our hypothesis: when the detector is exposed to sub-bandgap illumination, the measured signal of persistence, is worse immediately after saturation, but drops sooner and lower than the signal when there is no sub-bandgap illumination. The differences are small, but the result is repeatable. Further, if we increase the temperature of the soldering iron, even by 10-15K, the curve is significantly \emph{higher} than the non-illuminated control, indicating that the detector may be actually seeing in-band photons from a 350K thermal source.

From these results, we conclude that the infrared radiation did interact with the traps in the detector. By implementing similar changes to the setup as outlined in \S\ref{sec:cic_mitigate}, namely a using higher-flux source and/or a source emitting over a narrower wavelength range centered at a shorter wavelength, we may see a more significant improvement in the persistence decay curve. 

\section{CONCLUSIONS AND FUTURE WORK}
We have characterized the e2v CCD97 EMCCD as it operates in the N\"uv\"u HN\"u 512 camera, and found its levels of dark current and clock-induced charge to be equivalent to or better (lower) than reported by N\"uv\"u Cameras. We further began experimenting with the effect of sub-bandgap illumination on the detector in this camera housing to explore the possibility of lowering the combined dark current and CIC background floor even lower, which would in turn increase the efficiency of observations of extremely faint flux targets, such as directly imaged Earth analogs. 

As the results of the initial sub-bandgap experiments were either inconclusive, or showing effects of small magnitudes, we will continue the tests by improving the infrared source of sub-bandgap photons. We will work with sources emitting at shorter wavelengths, over narrower wavebands, and with much higher photon rates. The underlying physics of traps and the results so far from the persistence decay curve experiment make this approach a promising method for improving the performance of the camera in astrophysical applications. 

In the future, once characterizations and mitigation experiments are completed both on this HN\"u 512 and the larger-format HN\"u 1024, we will demonstrate the detector on-sky on the Apache Point Observatory Goddard Integral Field Spectrograph, which will further support the case for its use on space missions. The photon counting EMCCD continues to be a promising technology for direct imaging spectroscopy of exoplanets. 

\acknowledgments 
We would like to thank Brad Greeley for the gracious loan of the WFC-3 CASTLE integrating sphere. We also gratefully acknowledge the contributions to our program by Dr. Simon Tulloch at the University of Sheffield, including useful conversations, literature recommendations and providing software examples for EMCCD data analysis. The authors would like to acknowledge internal support provided by the Goddard Space Flight Center, and support from the NASA APRA and EPSCoR programs.  Finally, the authors would like to acknowledge the significant efforts, insights, and inspirations to this research by the late Bruce Woodgate. 


\bibliographystyle{spiebib2}

\end{document}

%% file: mymacros.tex
\def\sun{\hbox{$\odot$}}
\def\earth{\hbox{$\oplus$}}
\def\lesssim{\mathrel{\hbox{\rlap{\hbox{\lower4pt\hbox{$\sim$}}}\hbox{$<$}}}}
\def\gtrsim{\mathrel{\hbox{\rlap{\hbox{\lower4pt\hbox{$\sim$}}}\hbox{$>$}}}}

\def\arcdeg{\hbox{$^\circ$}}

\def\farcm{\hbox{.\kern -0.7ex\raisebox{.9ex}{\scriptsize$\prime$}}}
\def\farcs{\hbox{\kern 0.13ex.\kern -0.95ex\raisebox{.9ex}{\scriptsize$\prime\prime$}\kern -0.1ex}}

\def\apjl{ApJ}
\def\apjs{ApJS}

\def\aap{A\&A}

\def\eps@scaling{1.0}%
\newcommand\epsscale[1]{\gdef\eps@scaling{#1}}%
\newcommand\plotone[1]{%
 \typeout{Plotone included the file #1}
 \centering
 \leavevmode
 \includegraphics[width={\eps@scaling\columnwidth}]{#1}%
}%
\newcommand\plottwo[2]{{%
 \typeout{Plottwo included the files #1 #2}
 \centering
 \leavevmode
 \columnwidth=.48\columnwidth
 \includegraphics[width={\eps@scaling\columnwidth}]{#1}%
 \hfil
 \includegraphics[width={\eps@scaling\columnwidth}]{#2}%
}}%


%% file: ms.bbl
\begin{thebibliography}{10}

\bibitem{Shaklan2013}
{Shaklan}, S., {Levine}, M., {Foote}, M., {Rodgers}, M., {Underhill}, M.,
  {Marchen}, L., and {Klein}, D., ``{The AFTA Coronagraph Instrument},'' in
  [{\em Proc. SPIE}{\nolinebreak\hspace{0.1em}]},   {\bf 8864} (September
  2013).

\bibitem{Postman2010}
{Postman}, M., {Brown}, T., {Sembach}, K., {Giavalisco}, M., {Traub}, W.,
  {Stapelfeldt}, K., {Calzetti}, D., {Oegerle}, W., {Rich}, R.~M., {Stahl},
  H.~P., {Tumlinson}, J., {Mountain}, M., {Soummer}, R., and {Hyde}, T.,
  ``{Science drivers and requirements for an Advanced Technology Large Aperture
  Space Telescope (ATLAST): implications for technology development and
  synergies with other future facilities},'' in [{\em Proc.
  SPIE}{\nolinebreak\hspace{0.1em}]},   {\bf 7731} (July 2010).

\bibitem{McElwain2013}
{McElwain}, M.~W., {Perrin}, M.~D., {Gong}, Q., {Wilkins}, A.~N.,
  {Staplefeldt}, K.~R., {Woodgate}, B.~E., {Brandt}, T.~D., {Heap}, S.~R.,
  {Hilton}, G.~M., {Kruk}, J.~W., {Moody}, D., and {Trauger}, J., ``{PISCES: an
  integral field spectrograph to advance high contrast imaging technologies},''
  in [{\em SPIE Conference Series}{\nolinebreak\hspace{0.1em}]},   {\bf 8864}
  (2013).

\bibitem{Trauger2007}
{Trauger}, J., {Give'on}, A., {Gordon}, B., {Kern}, B., {Kuhnert}, A., {Moody},
  D., {Niessner}, A., {Shi}, F., {Wilson}, D., and {Burrows}, C., ``{Laboratory
  demonstrations of high-contrast imaging for space coronagraphy},'' in [{\em
  Proc. SPIE}{\nolinebreak\hspace{0.1em}]},   {\bf 6693} (Sept. 2007).

\bibitem{Levine2009}
{Levine}, M., {Soummer}, R., {Arenberg}, J., {Belikov}, R., {Bierden}, P.,
  {Boccaletti}, A., {Brown}, R., {Burrows}, A., {Burrows}, C., {Cady}, E.,
  {Cash}, W., {Clampin}, M., {Cossapakis}, C., {Crossfield}, I., {Dewell}, L.,
  {Egerman}, R., {Fergusson}, H., {Ge}, J., {Give'On}, A., {Guyon}, O., {Heap},
  S., {Hyde}, T., {Jaroux}, B., {Jasdin}, J., {Kasting}, J., {Kenworthy}, M.,
  {Kilston}, S., {Klavins}, A., {Krist}, J., {Kuchner}, M., {Lane}, B.,
  {Lillie}, C., {Lyon}, R., {Lloyd}, J., {Lo}, A., {Lowrance}, P.~J.,
  {Macintosh}, P.~J., {McCully}, S., {Marley}, M., {Marois}, C., {Matthews},
  G., {Mawet}, D., {Mazin}, B., {Mosier}, G., {Noecker}, C., {Pueyo}, L.,
  {Oppenheimer}, B.~R., {Pedreiro}, N., {Postman}, M., {Roberge}, A.,
  {Ridgeway}, S., {Schneider}, {Schneider}, J., {Serabyn}, G., {Shaklan}, S.,
  {Shao}, M., {Sivaramakrishman}, A., {Spergel}, D., {Stapelfeldt}, K.,
  {Tamura}, M., {Tenerelli}, D., {Tolls}, V., {Traub}, W., {Trauger}, J.,
  {Vanderbei}, R.~J., and {Wynn}, J., ``{Overview of Technologies for Direct
  Optical Imaging of Exoplanets},'' in [{\em astro2010: The Astronomy and
  Astrophysics Decadal Survey}{\nolinebreak\hspace{0.1em}]},  {\em Astronomy}
  (2010).

\bibitem{Hynecek2003}
Hynecek, J. and Nishiwaki, T., ``{Excess noise and other important
  characteristics of low light level imaging using charge multiplying CCDs},''
  {\em IEEE Transactions on Electron Devices}~{\bf 50},  239--245 (Jan. 2003).

\bibitem{Wen2006}
{Wen}, Y., {Rauscher}, B.~J., {Baker}, R.~G., {Clampin}, M.~C., {Fochie}, P.,
  {Heap}, S.~R., {Hilton}, G., {Jorden}, P., {Linder}, D., {Mott}, B., {Pool},
  P., {Waczynski}, A., and {Woodgate}, B., ``{Individual photon counting using
  e2v L3 CCDs for low background astronomical spectroscopy},'' in [{\em SPIE
  Conference Series}{\nolinebreak\hspace{0.1em}]},   {\bf 6276} (July 2006).

\bibitem{Tulloch2008}
Tulloch, S., ``{Modeling the suitability of EMCCDs for spectroscopic
  applications},'' in [{\em High Energy}{\nolinebreak\hspace{0.1em}]},  Dorn,
  D.~A. and Holland, A.~D., eds.,  {\bf 7021},  70212C--70212C--10 (Aug. 2008).

\bibitem{MacKay2004}
{MacKay}, C., {Basland}, A., and {Bridgeland}, M.

\bibitem{Ives2008}
{Ives}, D., {Bezawada}, N., {Dhillon}, V., and {Marsh}, T., ``{ULTRASPEC: an
  electron multiplication CCD camera for very low light level high speed
  astronomical spectrometry},'' in [{\em SPIE Conference
  Series}{\nolinebreak\hspace{0.1em}]},   {\bf 7021} (Aug. 2008).

\bibitem{Tulloch2010}
{Tulloch}, S.~M., ``Optimisation of an emccd,'' {\em Munich dfa}  (2010).

\bibitem{Daigle2009}
{Daigle}, O., {Carignan}, C., {Gach}, J.-L., {Guillaume}, C., {Lessard}, S.,
  {Fortin}, C.-A., and {Blais-Ouellette}, S., ``{Extreme Faint Flux Imaging
  with an EMCCD},'' {\em PASP}~{\bf 121},  866--884 (Aug. 2009).

\bibitem{Daigle2010}
{Daigle}, O., {Quirion}, P.-O., and {Lessard}, S., ``{The darkest EMCCD
  ever},'' in [{\em SPIE Conference Series}{\nolinebreak\hspace{0.1em}]},
  {\bf 7742} (July 2010).

\bibitem{Daigle2012}
{Daigle}, O., {Djazovski}, O., {Laurin}, D., {Doyon}, R., and {Artigau},
  {\'E}., ``{Characterization results of EMCCDs for extreme low-light
  imaging},'' in [{\em Proc. SPIE}{\nolinebreak\hspace{0.1em}]},   {\bf 8453}
  (July 2012).

\bibitem{RobbinsHadwen2003}
Robbins, M. and Hadwen, B., ``The noise performance of electron multiplying
  charge-coupled devices,'' {\em Electron Devices, IEEE Transactions on}~{\bf
  50},  1227--1232 (May 2003).

\bibitem{MacKay2012}
{Mackay}, C., {Weller}, K., and {Suess}, F., ``{Photon counting EMCCDs: new
  opportunities for high time resolution astrophysics},'' in [{\em SPIE
  Conference Series}{\nolinebreak\hspace{0.1em}]},   {\bf 8453} (July 2012).

\bibitem{Heap2008}
{Heap}, S.~R., {Lindler}, D., and {Lyon}, R., ``{Detecting biomarkers in
  exoplanetary atmospheres with a Terrestrial Planet Finder},'' in [{\em Proc.
  SPIE}{\nolinebreak\hspace{0.1em}]},   {\bf 7010} (Aug. 2008).

\bibitem{Michaelis2013b}
Michaelis, H., Behnke, T., Mottola, S., Krimlowski, A., Borgs, B., Holland, A.,
  and Schmid, M., ``{ Investigations on performance of Electron Multiplied CCD
  detectors (EMCCDs) after radiation for observation of low light star-like
  objects in scientific space missions },'' {\em Proc. SPIE}~{\bf 8889} (2013).

\bibitem{Rest2002}
Rest, A., M{\"u}ndermann, L., Widenhorn, R., Bodegom, E., and McGlinn, T.~C.,
  ``Residual images in charged-coupled device detectors,'' {\em Review of
  Scientific Instruments}~{\bf 73}(5) (2002).

\end{thebibliography}
